\documentstyle[12pt,twoside]{article}

\begin{document}
\begin{center}
\vspace{2.5cm} {\bf GRAVITOMAGNETIC EFFECTS IN A CONDUCTOR\\ IN AN
APPLIED MAGNETIC FIELD}\\ \vspace{2cm} B.J.
Ahmedov\footnote{E-mail: ahmedov@astrin.uzsci.net}\\ {\em Ulugh
Beg Astronomical Institute and Institute of Nuclear Physics\\
Ulughbek, Tashkent 702132, Uzbekistan\\ and\\ The Abdus Salam
International Centre for Theoretical Physics,
 Trieste, Italy}\\
\bigskip
and\\
\bigskip
M. Karim\\ {\em Department of Physics, St John Fisher College,
Rochester, NY 14618, USA}
\end{center}
\vspace{1cm}
\centerline{\bf Abstract}
\baselineskip=18pt
\medskip

The electromagnetic measurements of general relativistic gravitomagnetic
effects which can be performed within a conductor embedded in the
space-time of slow rotating gravitational object in the presence of
magnetic field are proposed.

\newpage

The general relativistic electromagnetic effects arising from
the gravitomagnetic field in noncurrent carrying (super-)conductors
with no applied magnetic field present
have been discussed by several authors
(see, for review, \cite{ignaz}).
However, the general-relativistic effects can be amplified by the
interplay between gravitomagnetic
field and either electric current or magnetic field
and in this respect, we discuss here a  new
test of gravitomagnetic field of Earth  by using conductors
embedded in an external magnetic field while in the previous paper~\cite{gg}
we have already shown that the interaction between the gravitomagnetic
field and electric current can lead to the galvanogravitomagnetic effect.

Space-time outside a spherically symmetric mass $M$ with angular
momentum $a$ is described by the Kerr metric. This differs from the
Schwarzschild solution for a static body by having non-diagonal
terms, which imply a local inertial frame to be rotating with
respect to the distant stars at infinity with
the Lense-Thirring angular velocity~\cite{bahram}
$\omega (r)\equiv {2aM}/{r^3}$.
Then the metric of the reference frame corotating with the slowly rotating
gravitational object with mass $M$ (in the linear angular momentum $a$
approximation) is
\begin{equation}
ds^2=-N^2
c^2dt^2+N^{-2}dr^2+r^2d\theta ^2+r^2\sin
{}^2\theta d\varphi ^2+
2{\bar{\omega}} r^2\sin {}^2\theta dtd\varphi ,
\label{eq:corot}
\end{equation}
where $N\equiv\left({1-{2M}/{r}}\right)^{1/2}$,
${\bar{\omega}} \equiv \Omega -\omega (r)$,
$\Omega$ is angular velocity of rotation of gravitational object with
respect to the distant stars.

Suppose that the material relations between inductions and fields
have linear character
\begin{eqnarray}
\label{relat}
H_{\alpha\beta}=\frac{1}{\mu}F_{\alpha\beta}+\frac{1-\epsilon\mu}{\mu}
(u_\alpha F_{\sigma\beta} -u_\beta F_{\sigma\alpha})u^\sigma , \nonumber\\
F_{\alpha\beta}=\mu H_{\alpha\beta}+\frac{\epsilon\mu -1}{\epsilon}
(u_\alpha H_{\sigma\beta} -u_\beta H_{\sigma\alpha})u^\sigma
\end{eqnarray}
 and the general relativistic  Ohm's law for conduction current
$\jmath^\alpha$ is
\begin{equation}
\frac{j_\alpha}{\lambda}=F_{\alpha\beta}u^{\beta}-R_H (F_
{\alpha\sigma}+F_{\rho\sigma}u^\rho u_\alpha )j^\sigma +
R_{gg}j^\beta A_{\alpha\beta}+{\Lambda^{-1/2}}
\stackrel{\perp}{\nabla}_\alpha(\Lambda^{1/2}\mu_e),
\label{ohm}
\end{equation}
here $F_{\alpha\beta}$ and $H_{\alpha\beta}$
are the tensors of electromagnetic field and induction, respectively,
$\epsilon$ and $\mu$ are the parameters for the conductor,
$\mu_e$ is the electrochemical potential per unit charge,
$R_{gg}={2mc}/{ne^2}$ is the parameter for the conductor
called as galvano-gravitomagnetic one, $n$ is the concentration
of the conduction electrons,
obviously $\lambda$ is the electrical conductivity, $R_H$ is the Hall
constant, $u_\alpha$ is the four-velocity
of the conductor, $w_\alpha =u_{\alpha ;\beta}u^\beta$ is the
absolute acceleration,
$A_{\beta\alpha}=u_{\alpha ,\beta}+u_{[\beta}w_{\alpha ]}$
is the relativistic rate of rotation of the conductor,
$\stackrel{\perp}{\nabla}_\alpha$ denotes the spatial part of
covariant derivative and $[\quad ]$ represents
antisymmetrization.
 The gravitational field is assumed to be
stationary that is space-time metric $g_{\alpha\beta}$ admits a
timelike Killing vector $\xi^\alpha_{(t)}$
that is
${\it\L}_{\xi_t} g_{\alpha\beta}=0$ ({\it\L}$_{\xi_t}$ denotes the Lie
derivative with respect to $\xi^\alpha_{(t)}$,
$\Lambda =-\xi^\alpha_{(t)}\xi_{(t)\alpha}$).

\newpage

Then the general relativistic expression for the charge distribution
inside conductor~\cite{aa94}

\begin{eqnarray}
&&\rho_0 = \frac{\epsilon\mu R_H}{c}\jmath^2+\frac{1}{4\pi}
    \{(\frac{\epsilon}
    {\lambda}\jmath^\alpha)_{;\alpha}+[\epsilon^2\mu
    R_H(\frac{1}{\lambda}\jmath^2+
    {\Lambda^{-1/2}}\jmath^\nu\stackrel{\perp}{\nabla}_\nu
    (\Lambda^{1/2}\mu_e))u^\alpha ]_{;\alpha}
\nonumber\\
&&-\epsilon R_{gg}A_{\alpha\beta}
w^\alpha\jmath^\beta+
g^{\alpha\beta}(\epsilon R_{gg}\jmath^\nu
A_{\alpha\nu})_{;\beta}-
\frac{\epsilon}{\lambda}w^\alpha\jmath_\alpha -
\epsilon w^\alpha{\Lambda^{-1/2}}
\stackrel{\perp}{\nabla}_\alpha(\Lambda^{1/2}\mu_e)\nonumber\\
&&+ g^{\alpha\beta}(\epsilon {\Lambda^{-1/2}}
\stackrel{\perp}{\nabla}_\alpha(\Lambda^{1/2}\mu_e))_{;\beta}
+H^{\alpha\beta}[A_{\beta\alpha}+\epsilon\mu R_Hw_\alpha\jmath_\beta
+(\epsilon\mu R_H\jmath_\alpha)_{,\beta}]\}
\end{eqnarray}
can be derived from the general relativisic Maxwell equations
\begin{equation}
e^{\alpha\beta\mu\nu} F_{\beta\mu ,\nu}= 0,\quad
{H^{\alpha\beta}}_ {;\beta}= \frac{4\pi}{c}J^\alpha,\quad
J^\alpha\equiv c\rho_0 u^\alpha +\hat\jmath^\alpha.
\label{eq:max}
\end{equation}
by using material relationships (\ref{relat}) and (\ref{ohm}).

The charge density $\rho_0$ inside a conductor which has
no conduction current $\jmath =0$ but
embedded in an external magnetic field  ${\mathbf B}$ is
\begin{equation}
\rho _0=\frac{1}{4\pi}\{\epsilon Aw^2-(\epsilon Aw^\alpha)_{;\alpha}+
F^{\alpha\beta}A_{\beta\alpha}\}
\label{eq:charge}
\end{equation}
and has two contributions: the first one is due to the absolute
acceleration $w_\alpha$ and
second one is due to the relativistic rate of rotation of the conductor
$A_{\beta\alpha}$ and can be
adjusted and amplified by the magnetic field penetrating
inside the conductor. Here
$F_{\alpha\beta}=2\tau_{[\alpha} E_{\beta ]} +
\eta_{\alpha\beta\mu\nu}\tau^\mu B^\nu$ is the electromagnetic field
tensor, $A$ is the parameter for the conductor,
$\tau^\alpha$ is four-velocity of observer, $E^{\alpha}$ and
$B^\alpha$ are the electric and magnetic fields measured by observer.

In our approximation the charge density, inside a conductor at rest in
the orbiting station~(\ref{eq:corot}), is

\begin{equation}
\rho_0=-\frac{1}{2\pi}\left\{F^{23}A_{23}+F^{13}A_{13}\right\}.
\end{equation}
We do not consider the charge redistribution arising from the absolute
acceleration of the conductor since it does not depend on electromagnetic
field characteristics.

If the electromagnetic field tensor components are
\begin{equation}
F^{31}=\frac{NB^\theta}{r\sin\theta}, \qquad
F^{23}=\frac{B^r}{r^2\sin\theta}
\end{equation}

and the nonvanishing components of the relativistic rate of rotation
have form

\begin{equation}
A_{13}=\frac{\Omega r+\omega r /2}{cN}
\sin {}^2\theta ,
\qquad A_{23}=\frac{\bar{\omega}r^2}
{cN}\sin\theta\cos\theta ,
\label{eq:rot}
\end{equation}

then in our approximation, the space charge density inside the
conductor at rest in the frame of reference
~(\ref{eq:corot}) is
\begin{equation}
\rho_0=\frac{\Omega}{2\pi c}\left[B^{\theta}\sin\theta
- \frac{B^{r} \cos\theta}{N} \right]
+\frac{\omega}{4\pi c} \left[\frac{2B^{r}\cos\theta}{N} +
B^{\theta}\sin\theta \right] \ ,
\label{eq:rho}
\end{equation}
where the magnetic field components are measured by zero angular
momentum observers with four-velocity
$\tau_\alpha\equiv\{-N,0,0,0\}$.

The first term in the right hand side of equation (\ref{eq:rho})
results from angular velocity $\Omega$ and last one is due to the
gravitomagnetic field of the Earth and has pure general relativistic
nature.

On Earth, the angular velocity of the conductor is given
by~\cite{ignaz} ${\mathbf\Omega}_{cond}={\mathbf\Omega}-
{\mathbf\Omega}_{Th}-{\mathbf\Omega}_S-{\mathbf\omega}$,
where ${\mathbf\Omega}_{Th}$ and ${\mathbf\Omega}_S$
are the contributions of the Thomas precession arising
{}from non-gravitational forces and of the de Sitter or
geodetic precession. In order to measure ${\mathbf\omega}$
 one should measure ${\mathbf\Omega}_{cond}$ and then substract
{}from it the independently measured value of ${\mathbf\Omega}$
with Very Long Baseline Intererometry~\cite{koval} and the
contributions due to the Thomas and de Sitter precession.

In contrast to ({\ref{eq:rho}), for a superconductor embedded
in the gravitational field
(\ref{eq:corot}) the space charge density $\rho_0 (sc)=0$, that is
according to the solutions of the general-relativistic Maxwell
equations and London equations, the magnetic field penetrating
superconductor is proportional to ${\bar\omega}$ and consequently
the charge density is at least of order of ${\bar\omega}^2$.
Therefore, if the temperature T is increased then in the point
of the phase transition $T=T_c$ the applied magnetic field
penetrates inside the sample and induces a nonvanishing
charge density with the corresponding flow of charges.

For the Earth with mass $M =0.44cm$ and radius
$R\approx 6.37\times 10^8cm$,
$\omega (r)=\frac{4M}{5R}\Omega\approx 0.6\times 10^{-9}
\Omega$ near the surface. If the value of applied magnetic
field around conductor
is $10^3 G$ and the relaxation time $t_{rel}=10^{-8}s$
then one can find a typical value of charge exchange current
arising from gravitomagnetic Lense-Thirring frequency is of
order $10^{-14}A$ which is
within capacity of modern technical measurements.
However, in the present case, there are serious problems
arising from environment and the design of proposed experiment
is under consideration.

\vspace*{1cm} \noindent \section*{Acknowledgments}

BA acknowledges the hospitality at the Max-Planck-Institute
f$\ddot{u}$r Gravitationsphysik, Golm and the Abdus Salam
International Centre for Theoretical Physics, Trieste, during his
visit in autumn 1999 and the financial support from
Friedrich-Shiller Universitat, Jena, for his participation at the
Journees Relativiste 1999.

\end{document}